\documentclass[
]{ceurart}

\usepackage{listings}
\usepackage{latexsym}
\usepackage{amssymb}
\usepackage{amsmath}
\usepackage{amsthm}
\usepackage{booktabs}
\usepackage{enumitem}
\usepackage{graphicx}
\usepackage{color}
\usepackage{rotating}
\usepackage{booktabs}
\usepackage{tabularx}
\usepackage{array}
\usepackage[utf8]{inputenc}
\usepackage[greek,english]{babel}
\usepackage{amsmath}
\usepackage{placeins}
\usepackage{float} 
\usepackage{amsmath,bm} 
\usepackage{caption}
\usepackage{subcaption}
\usepackage{longtable}
\usepackage{booktabs}
\usepackage{wrapfig}
\begin{document}

\copyrightyear{2025}
\copyrightclause{Copyright for this paper by its authors.
  Use permitted under Creative Commons License Attribution 4.0
  International (CC BY 4.0).}

\conference{Identity-Aware AI workshop at 28th European Conference on Artificial Intelligence,
  October 25, 2025, Bologna, Italy}

\title{Visual Stereotypes of Autism Spectrum in Janus-Pro-7B, DALL-E, Stable Diffusion, SDXL, FLUX, and Midjourney}

\author[1]{Maciej Wodziński}[%
orcid=0000-0001-6347-5634,
email=maciek.wodzinski@gmail.com,
]
\cormark[1]

\author[1]{Marcin Rządeczka}[%
orcid=0000-0002-8315-1650,
email=marcin.rzadeczka@umcs.pl,
]

\author[5]{Anastazja Szuła}[%
orcid=0000-0002-2777-794X,
email=anastazja.szula@gmail.com,
]

\author[2,3,4]{Kacper Dudzic}[%
orcid=0009-0003-7849-6931,
email=kacper.dudzic@ideas.edu.pl,
]

\author[1,2,5]{Marcin Moskalewicz}[%
orcid=0000-0002-4270-7026,
email=marcin.moskalewicz@ideas.edu.pl,
]

\address[1]{Maria Curie-Skłodowska University, Plac Marii Curie-Skłodowskiej 4, 20-031 Lublin, Poland}
\address[2]{IDEAS Research Institute, Królewska 27, 00-060 Warsaw, Poland}
\address[3]{Adam Mickiewicz University, Uniwersytetu Poznańskiego 4, 61-614 Poznań, Poland}
\address[4]{AMU Center for Artificial Intelligence, Uniwersytetu Poznańskiego 4, 61-614 Poznań, Poland}
\address[5]{Poznań University of Medical Sciences, Rokietnicka 7, 60-806 Poznań, Poland}

\cortext[1]{Corresponding author.}

\begin{abstract}
Avoiding systemic discrimination of neurodiverse individuals is an ongoing challenge in training AI models, which often propagate negative stereotypes. This study examined whether six text-to-image models (Janus-Pro-7B VL2 vs. VL3, DALL-E 3 v. April 2024 vs. August 2025, Stable Diffusion v. 1.6 vs. 3.5, SDXL v. April 2024 vs. FLUX.1 Pro, and Midjourney v. 5.1 vs. 7) perpetuate non-rational beliefs regarding autism by comparing images generated in 2024-2025 with controls. 53 prompts aimed at neutrally visualizing concrete objects and abstract concepts related to autism were used against 53 controls (baseline total N=302, follow-up experimental 280 images plus 265 controls). Expert assessment measuring the presence of common autism-related stereotypes employed a framework of 10 deductive codes followed by statistical analysis. Autistic individuals were depicted with striking homogeneity in skin color (white), gender (male), and age (young), often engaged in solitary activities, interacting with objects rather than people, and exhibiting stereotypical emotional expressions such as sadness, anger, or emotional flatness. In contrast, the images of neurotypical individuals were more diverse and lacked such traits. We found significant differences between the models; however, with a moderate effect size (baseline $\eta^2 = 0.05$ and follow-up $\eta^2 = 0.08$), and no differences between baseline and follow-up summary values, with the ratio of stereotypical themes to the number of images similar across all models. The control prompts showed a significantly lower degree of stereotyping with large size effects (DALL·E 3 $\eta^2$  = 0.39; Midjourney $\eta^2$  = 0.41; FLUX  $\eta^2$  = 0.20; Stable Diffusion $\eta^2$  = 0.34; DeepSeek-VL3  $\eta^2$  = 0.45), confirming the hidden biases of the models. In summary, despite improvements in the technical aspects of image generation, the level of reproduction of potentially harmful autism-related stereotypes remained largely unaffected. 
\end{abstract}


\maketitle

\vspace*{-20pt}

\section{Introduction}
\noindent \textbf{Stereotypes in text-to-text and text-to-image models}\\
Analyses of AI cognitive biases and oversimplifications in their representations of various social phenomena play a significant role in AI ethics and fairness \citep{Cao2022} \citep{Mattern2022}. To prevent the perpetuation of systemic discrimination, it is imperative that users of Large Language Models (LLMs) are cognizant of their inherent limitations and that developers can identify and rectify them. 

Previous research has demonstrated that many models reproduce gender, race, age \citep{Zekun2023}, or ethnic stereotypes, and that AI models underlying assistive technologies contain biased stereotypes \citep{Herold2022}. For example, LLMs associate Muslims with violence \citep{Abid2021}; even when a model is tasked with generating content pertaining to Arabic culture, it remains ‘contaminated’ by the elements characteristic of the West \citep{Naous2023}. Furthermore, some models exhibit biases towards the values of specific societies \citep{Johnson2022} or may be biased politically \cite{Almandil2019}. While the majority of research in this field has focused on text-to-text models, a few studies have examined models that generate images from textual prompts \citep{Lin2023}. This is especially relevant in terms of human identity. While the deeper aspects of identity that have to do with the sense of self strongly relate to language, the more superficial self-perception is typically mediated by appearances. Bianchi demonstrated that text-to-image generation models amplify demographic stereotypes \citep{Bianchi2022}, and Aldahoul highlighted the presence of racial and gender stereotypes in AI-generated faces across 6 races, 32 professions, and 2 genders, and additionally proposed some debiasing solutions \citep{AlDahoul2024}. When investigated LLMs were asked to depict an ‘attractive person’, they predominantly depicted white individuals. In contrast, when LLMs were asked to depict a 'poor person', they predominantly depicted black individuals. In a similar vein, LLMs depicted a ‘terrorist’ as a Middle Eastern man. Even when explicitly instructed to depict a ‘white terrorist,’ the models generated images of a bearded man who visually resembled a stereotypically Middle Eastern individual. LLMs also perpetuate stereotypes concerning race, gender, and religion \citep{Wang2023}. For example, when asked to show ‘people who are political elites’, they show mainly white males.  

Most recently, DeepSeek's Janus-Pro-7B was reported to outperform OpenAI's DALL-E 3 and Stability AI's Stable Diffusion in text-to-image generation benchmarks, particularly allegedly achieving an 80\% accuracy rate on the GenEval \citep{Ghosh2023} benchmark compared to DALL-E 3's 67\% and Stable Diffusion's 74\%. Its enhanced performance was attributed to improvements in training processes, including the integration of 72 million high-quality synthetic images balanced with real-world data, resulting in more stable and detailed image generation \citep{Chen2025} \citep{Ma2024}. 

These claims rely on specific evaluation metrics and datasets, which may not fully capture real-world performance across diverse prompts and creative tasks. At the moment, there is no publicly available information indicating whether DeepSeek's Janus-Pro-7B model has been evaluated for biases or stereotypes, including neurodiversity-related stereotypes of interest. The existing benchmarks, such as GenEval and DPG-Bench, primarily assess the models’ ability to follow text prompts and generate images accurately in the technical sense of the word. 

To address the issue of reproducing stereotypical beliefs, AI developers often use 'fairness protocols', which are top-down safeguards. These protocols result in the models either refusing to generate specific content (e.g., a description of a representative of a particular social group) or circumventing the issue by generating content on similar substitute topics. This method is only a secondary and temporary solution, as it does not address the fundamental issue of biased training datasets.  

This prospective longitudinal study examines the degree of harmful representations regarding socially prevalent (and, therefore, likely included in training datasets) stereotypes about the autism spectrum in text-to-image models in 2024-2025. \vspace{1em}

\noindent \textbf{Socially prevalent stereotypes about the autism spectrum condition}\\
The example of autism is pertinent for a number of reasons. Firstly, the topic is of significant social importance and sensitivity, impacting ca. 62/10 000 people in the global population \citep{Elsabbagh2012}. Secondly, it is becoming increasingly prominent in the public eye. Numerous, often detrimental, identity stereotypes and oversimplifications about autism have been created and disseminated, and have become deeply embedded in collective awareness. For instance, although there is some evidence suggesting the prevalence of autistic cognitive style among STEM/IT professionals, stereotypically identifying all people on the spectrum with the figure of a brilliant computer geek is unsubstantiated \citep{Silberman2016}. Thirdly, the topic is characterized by a high degree of cognitive uncertainty, both in the social and scientific spheres. A number of studies have highlighted the historical variability and social construction of the autism category \citep{Draaisma2009} \citep{Wodzin2022}. Consequently, numerous beliefs and stereotypes about autism operate unconsciously in social awareness as the so-called background knowledge, influencing the identities of autistic individuals \citep{Camus2024}. The pervasive belief that autism is invariably accompanied by suffering, that the source of this suffering is the condition itself and not social misunderstanding, and that autism is primarily diagnosed in children, boys, and white individuals often leads to the perpetuation of hurtful prejudice against autistic individuals, impeding their social functioning and access to diagnosis and appropriate therapy.

Consequently, the beliefs about autistic identity propagated by AI models significantly influence opinions in this field as the topic gains increasing popularity in various spheres of public life  \citep{Treweek2019}. A multitude of stereotypes and myths surrounding autism negatively impact the lives of individuals on the spectrum. Autism communities seek to challenge these stereotypes and hegemonic narratives, aiming to redefine autism as a distinct mode of functioning, resulting, among other factors, from the atypical structure of the nervous system. Consequently, they seek to challenge the perception of autism as a deficit and to depathologize it, thereby reducing social stigma \citep{Kim2024}. In this context, the growing role that AI models play in shaping public awareness of neurodiversity makes it increasingly important to control for cognitive biases and non-rational beliefs in the models' performance. 

Table~\ref{tab:codes} presents deductive codes representing stereotypes selected for this study based on the literature review and our previous research, along with their operationalized definitions and a brief explanation of their harmfulness. 

\section{Methods}
\textbf{Research protocol}\\
The research protocol involved generating images in two rounds, one year apart, except for Janus Pro, which was released only in early 2025, hence the gap between rounds was 4 months (total N=302 at baseline and N=280 in the follow-up), based on 53 distinct prompts, selected with the objective of visualizing, in a possibly neutral way, concrete objects and abstract concepts related to autism across five models. The follow-up aimed to determine whether the advances in the technical aspect of image generation led to a reduction in the degree of use of stereotypical motifs associated with autism. In addition, 53 control prompts were used in the follow-up to account for the randomness of stereotypization. 

DALL-E 3 \citep{Betker} (v. April 2024 and v. August 2025) is based on an undisclosed architecture. Stable Diffusion \citep{Rombach2022} (v. 1.6 medium at baseline and v. 3.5 medium in the follow-up) employs a latent diffusion model, which processes images in a compressed feature space and gradually refines the image from a random noise distribution through a series of learned reverse diffusion steps (that use stochastic processes to create images from initial noise). FLUX.1 Pro \citep{Chen2025} is built upon a hybrid architecture of multimodal and parallel diffusion transformer blocks, scaled to 12B parameters. Midjourney's (v. 5.1 at baseline and v. 7 in the follow-up) architecture is also unknown, while Janus-Pro (v. VL2 at baseline and v. VL3 in the follow-up) decouples visual encoding into separate pathways, while still utilizing a single unified transformer architecture for processing. The DALL-E 3, Stable Diffusion, and FLUX models were used through the dedicated Python API's provided by their developers; the Midjourney model was used through the GUI available on its official website, whereas the Janus-Pro model was run locally on a single NVIDIA A100 80GB GPU using the text-to-image generation script provided on its GitHub page. For all the models, the default provided inference settings were used.

The experimental prompts were engineered to ensure a neutral form without suggesting the use of specific symbols or themes. The issues covered were selected by a team of experts, including a person on the spectrum (hence: participatory prompts co-design with members of the autistic community) to take into account both the image of autistic people (individually and in groups) and various types of behavior, interactions, and everyday situations. However, the team also considered more abstract concepts, such as the visualization of the phenomenon of autism itself or emotions.  

Prompts were phrased to focus on lived experience rather than pathology, e.g., instead of deficit-based semantics, “difficulty caused by autism” - “difficulty faced by an autistic person”, not to imply normative judgments. Our assessment was based on group discussion and iterative testing, where interfaces allowed (e.g., models asked how they interpret phrases). While some prompts may seem redundant, it was a deliberate design choice to include closely semantically related prompts but with minor variance in phrasing (alternate sentence structure) to control for semantic bias (outputs skewed by lexical choices) and ensure that any stereotypes detected were attributable to the concept (e.g., autism), rather than prompt form. For the control group, we prepared a symmetrical set of prompts aimed at representing non-autistic individuals. It was done by either deleting the word “autistic” (e.g. “Create an image of an autistic person” - “Create an image of a person,”) or replacing the abstract concept of “autism” with “neurotypicality” (e.g. “Describe autism with an image” - “Describe neurotypicality with an image”, “visualize autism” - “visualize neurotypicality”). The control-prompt group isn’t fully neutral due to the concept of neurotypicality, but it fulfills the key function of controlling for autism-related content. This choice allows for a direct conceptual contrast between “autism” and “neurotypicality,” thereby strengthening the clarity and consistency of the control condition. Also, some visual biases (gender or age-related) may appear in the models independently of the concept of autism, and either represent training data or amplify autism-specific stereotypes, while some prompts (e.g., “at school”) may implicitly cue certain representations (e.g., children). We attempted to minimize such effects, but acknowledge that some prompts may prime certain representations (e.g., “school” often evokes children), which is a limitation. To indirectly control for false positives, we took into account the over-representation of children against the over-representation of white boys in those few prompts.

Each prompt was administered once to each model. However, the final number of images exceeds the number of prompts multiplied by the number of models because some models generated multiple alternative versions. When requested to generate multiple themes or objects, the models occasionally returned the results as a single image (e.g., split into three parts) and at other times as three separate images. Midjourney consistently generated four preliminary images. To avoid arbitrariness in the selection, all images generated by all models were included in the analysis, resulting in uneven baseline and follow-up samples. 
The default hyperparameter values of models were retained between the baseline and follow-up versions; for closed-source models (all except Janus), this was limited to the equivalence of the hyperparameter and model snapshot variable values. Reproducibility of the outputs was only checked ad hoc and not quantitatively assessed, which is a limitation.
The results were subjected to an expert assessment of independent coders via a framework of 10 deductive codes that represented common stereotypes contested by the autistic community, regarding their presence, judged by taking into consideration their spatial intensity on an image (see Table~\ref{tab:codes}). The presence of a stereotype was rated on a categorical scale, yes/no, while the overall level of stereotyping was determined by adding up the scores across the list of ten deductive codes, with each image subsequently rated on this 0-10 ordinal scale. This is referred to as the ‘degree of stereotyping'. The results were subjected to statistical analysis of inter-rater reliability and size effects. The full research protocol, including the comprehensive list of prompts, all generated images, the evaluation form, and the inter-rater reliability assessment, is attached as supplementary material, which can be downloaded here: \url{https://figshare.com/s/8caea1bd2c2910598b98}. \vspace{1em}  

\noindent \textbf{Community involvement}\\
The first author of this paper is an active autistic researcher and a parent of two autistic children, and another co-author is on the spectrum. It has now become generally accepted by the participatory research community that autistic people provide unique epistemic perspectives to the field of autism research. \vspace{1em}

\noindent \textbf{Data Analysis}\\
In the baseline data analysis (N=302), three subsequent pilot coding sessions on randomized samples of 20 from the dataset were conducted to improve inter-rater reliability using Cohen's kappa coefficient, accounting for agreement occurring by chance. We have improved the initial inter-rater reliability of 0.315 in the first pilot coding, through 0.698 in the second, to 0.93 in the third, mainly by redefining the operational qualitative descriptions of codes. In the follow-up rounds (experimental and control), a similar process of pilot coding was performed, resulting in an inter-rater reliability of 0.90. Hence, there were three subsequent coding sessions, both at baseline and in the follow-up; while definitions remained unchanged, the follow-up images differed; inter-rater reliability is provided independently for both baseline and follow-up. The final moderate k-values of 0.74 for baseline and 0.79 and 0.54 for follow-up represent the kappa paradox because the absolute agreement was > 0.8; while in the coding of controls, the presence of stereotypes was very low, hence there were a lot of 0 values accounting for the score (see research protocol for details). To calibrate the assessment framework and ensure its accuracy, these sessions included consensual adjustments to the qualitative evaluation grid based on feedback from the raters, which led to a progressive increase in inter-rater reliability. In effect, we refined the coding framework by increasing its specificity and sensitivity, taking into account those cases where there were divergencies between the evaluators, leading to both false-positive and false-negative results. For example, the “lonely” stereotype was assessed only when there were multiple people or contextual elements present in the image, e.g., when an individual was visually isolated from a group, placed apart in a playground. Such criteria were clearly specified in our coder guidelines to avoid over-attribution. To obtain unambiguous and non-fractional values, remaining minor differences between the two raters were solved by a third rater in a final meta-evaluation for all three sets of images. 

\begin{figure}[h!]
    \centering 
    
    \begin{subfigure}[t]{0.48\textwidth}
        \centering
        \includegraphics[width=\textwidth]{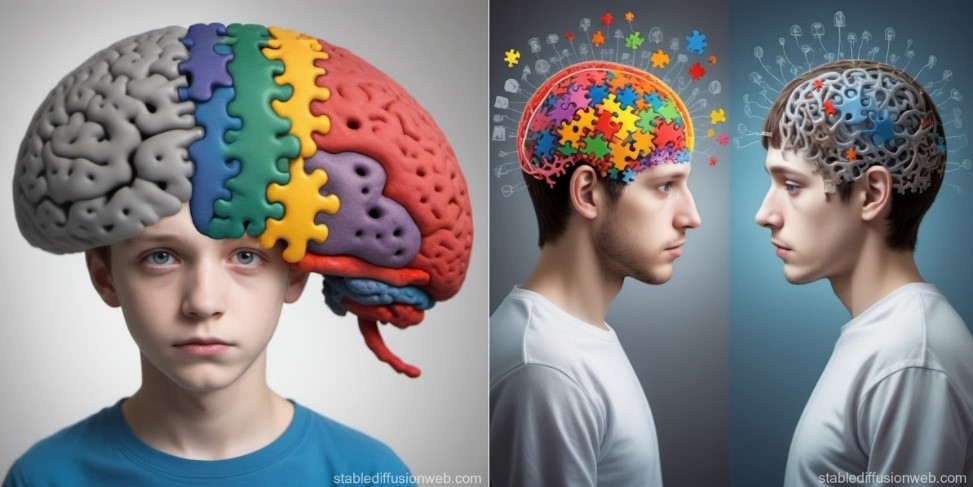}
        \caption{Two images combining many stereotypes. Prompts no. 15 (left) and 12 (right). Model: Stable Diffusion v. 1.6}
        \label{fig:SD_15}
    \end{subfigure}
    \hfill 
    \begin{subfigure}[t]{0.48\textwidth}
        \centering
        \includegraphics[width=\textwidth]{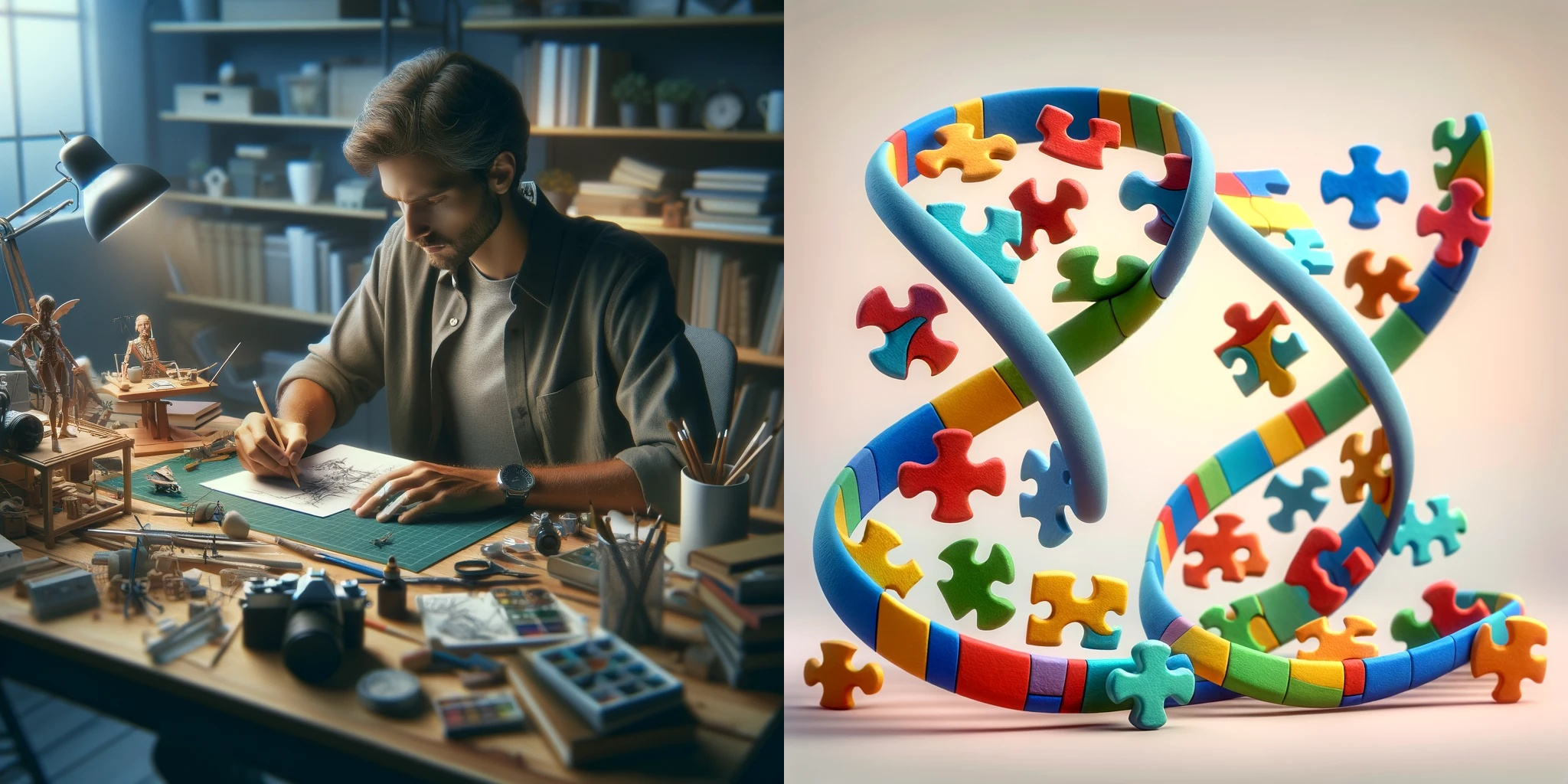}
        \caption{Examples of ambiguous images. Prompts no. 8 (left) and 38 (right). Model: DALL-E (April 2024)}
        \label{fig:GEEK}
    \end{subfigure}
    
    \caption{Comparison of stereotypical (a) and ambiguous (b) images from different models}
    \label{fig:side_by_side}
\end{figure}

Not all stereotypes were as readily apparent as those in Fig.~\ref{fig:SD_15}. Fig.~\ref{fig:GEEK} illustrates examples of ambiguous images for which expert raters were required to clarify the definitions of stereotypes (e.g., the concepts of a ‘nerd’ or blue color dominance) to achieve the desired level of agreement.

\begin{figure}[h!]
    \centering
    
    \begin{subfigure}[t]{0.48\textwidth}
        \centering
        \includegraphics[width=\textwidth]{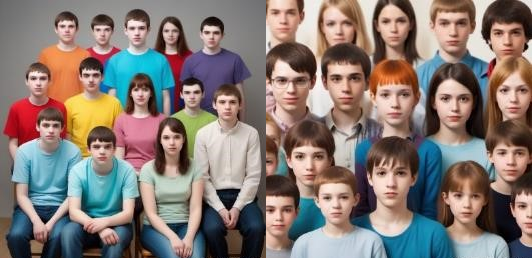}
        \caption{A homogenous group of people presenting stereotypical autistic characteristics. Prompt no. 11 (left) and 10 (right). Model: Stable Diffusion v. 1.6}
        \label{fig:group_of_people}
    \end{subfigure}
    \hfill 
    \begin{subfigure}[t]{0.48\textwidth}
        \centering
        \includegraphics[width=\textwidth]{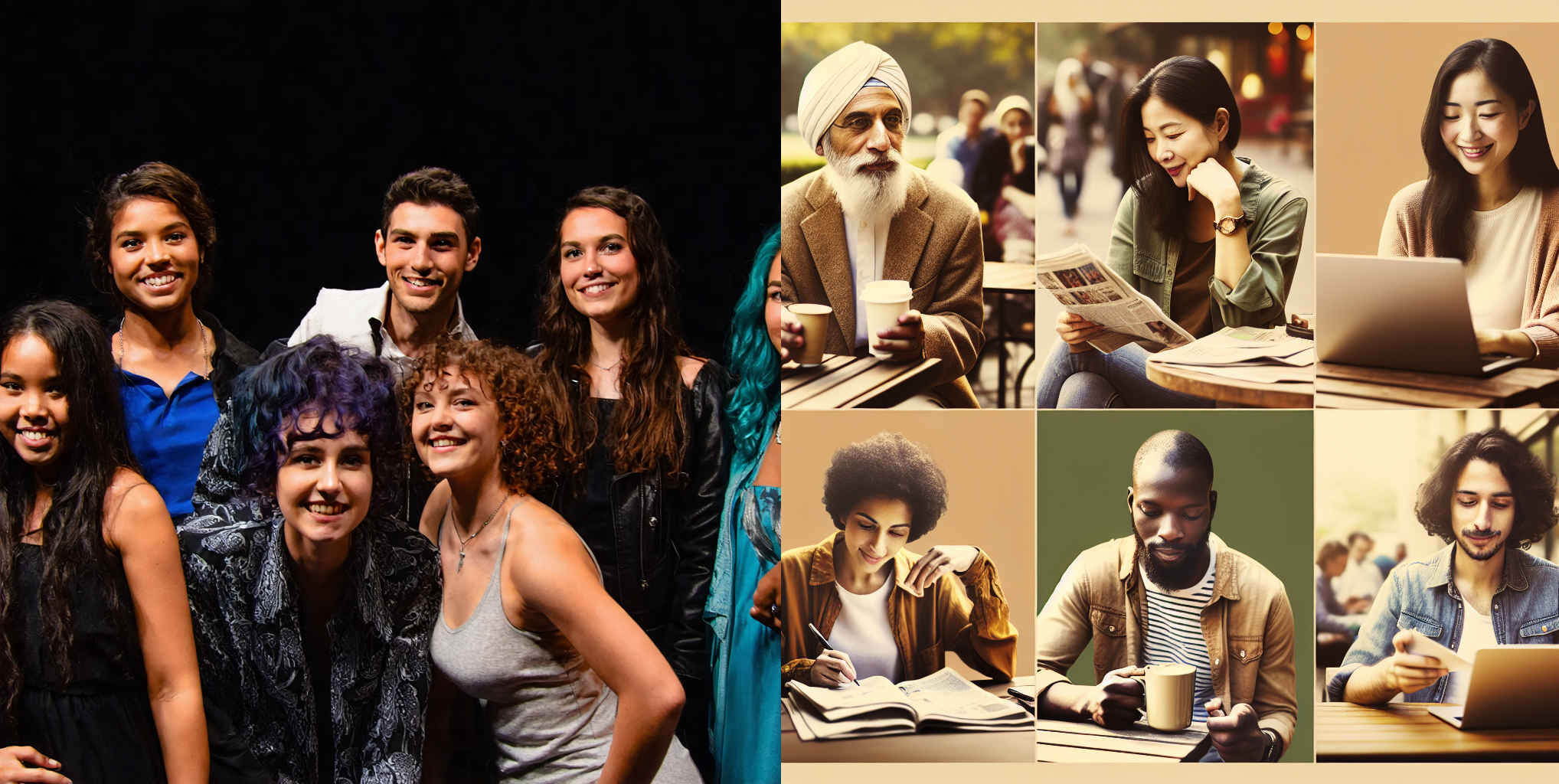}
        \caption{Control images showing more diverse groups of people. Prompt no. 11. Models: Stable Diffusion v. 3.5 (left) and DALL-E (August 2025)}
        \label{fig:controls_group_of_people}
    \end{subfigure}
    
    \caption{Comparison of AI-generated images for groups of people, showing stereotypical depictions (a) versus more diverse control images (b)}
    \label{fig:group_comparison}
\end{figure}

Images depicting groups of autistic individuals tend to present them in a highly homogeneous and uniform manner, with limited variation in characteristics such as gender, age, or skin color, which was not the case for the control group (see Fig.~\ref{fig:group_of_people} and Fig.~\ref{fig:controls_group_of_people}). To ascertain the presence of the white boy stereotype, three distinct codes had to be identified: white, including a greater proportion of individuals with white skin; child, including a greater number of children (with teenagers) than adults; and male, including a greater number of males than females.

\begin{figure}[h]
	\centering
	\includegraphics[width=7.5cm]{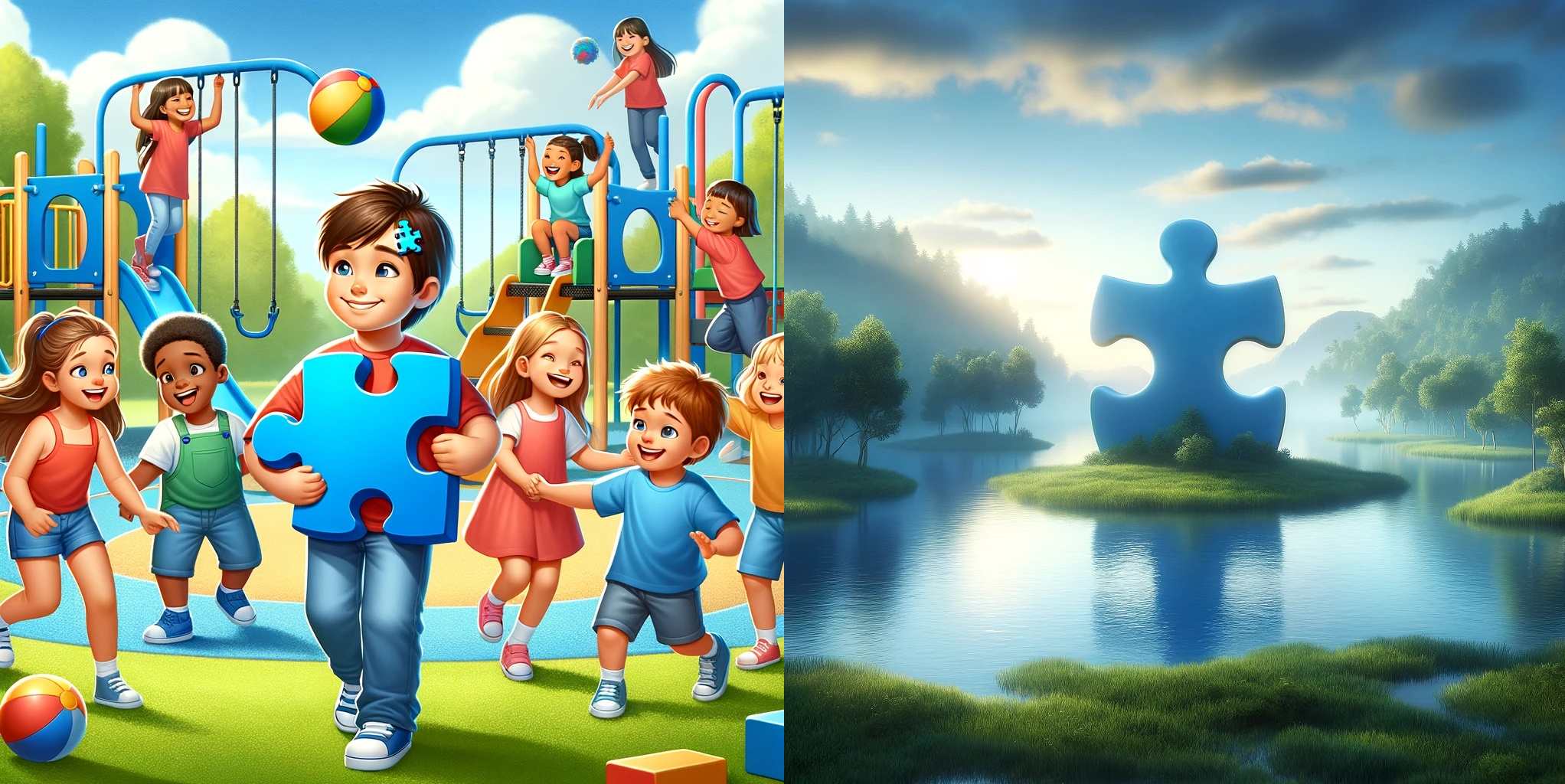}
	\caption{The prevalence of blue color and white boy themes. Prompt no. 51 (left) and 7 (right). Model: DALL-E (April 2024)}
	\label{fig:NIEB}
\end{figure}

The most frequently repeated stereotypical themes were the puzzle symbol and the blue color. The overwhelming majority of characters depicted were white boys. These three stereotypes represent a significant challenge for the autism community, which has been striving to combat them for years.  Consequently, the puzzle stereotype was operationalized more sensitively, with its presence being considered in any location within the image, including the background and edges. Similarly, the occurrence of a stereotypical association with the color blue was considered if this color appeared in the image more often than other colors (it did not have to constitute more than 50\% of the image area) or if blue was associated with a significant object, for example, located in the central, attention-grabbing part of the image (see Fig.~\ref{fig:NIEB}).

\vspace*{-20pt}

\section{Results}

\begin{wrapfigure}[25]{r}{0.4\textwidth}
    \centering
    \vspace{-10pt} 
    \includegraphics[scale=0.4]{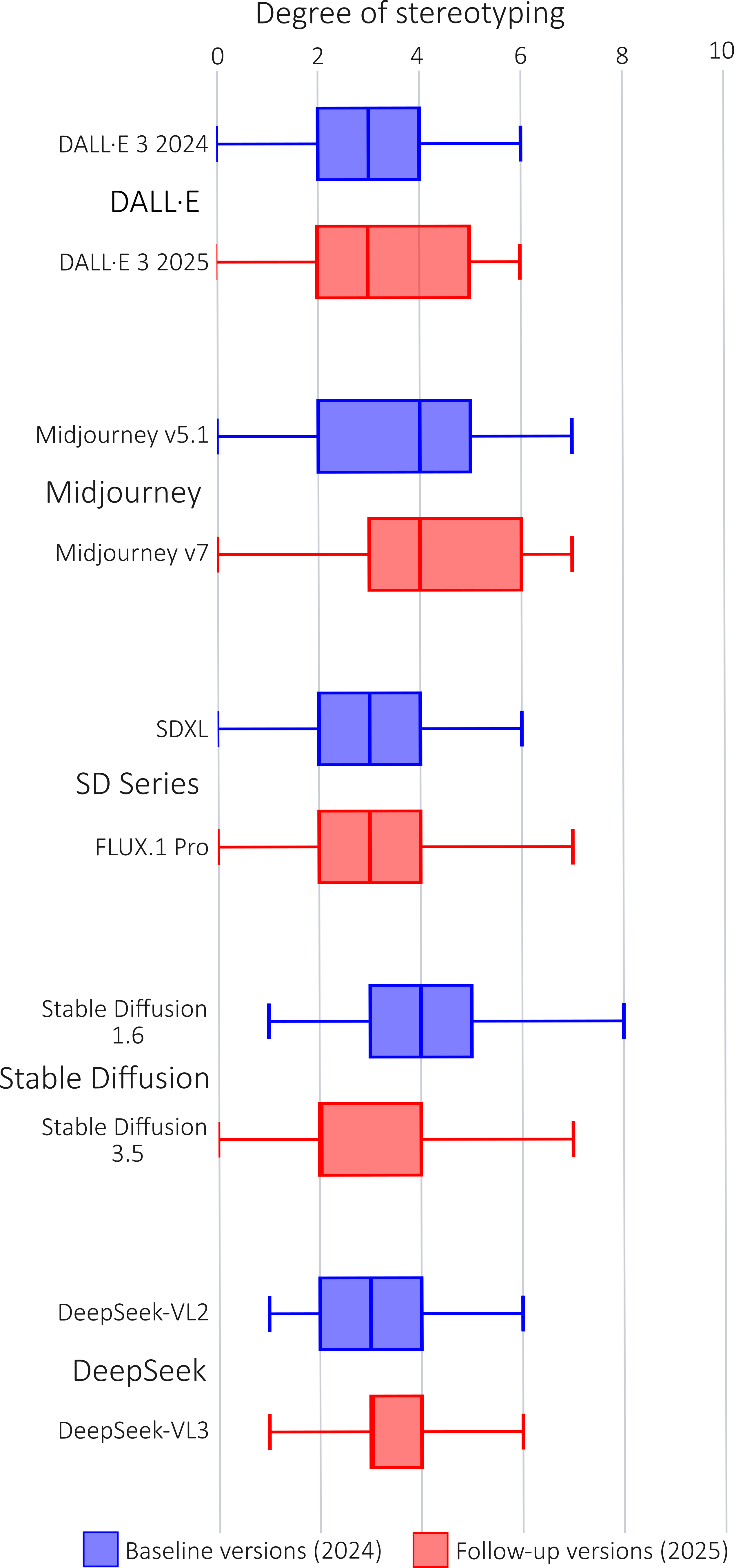}
    \caption{: Comparison of the distribution of the degree of stereotyping (0-10 scale) across five models at baseline and follow-up}
    \label{fig:10boxplots}
    \vspace{-15pt} 
\end{wrapfigure}

Distributions of the degree of stereotyping for all models differ significantly from normal. Testing (Kruskal-Wallis) indicated significant differences between the degree of autistic stereotyping between all models, however, with a moderate effect size (for baseline $\eta^2$ = 0.05, for the follow-up $\eta^2$ = 0.08). The highest degree of stereotyping was observed for Stable Diffusion (M/me 3.915/4.00) at baseline and for FLUX (M/me 4.151/4.00) in the follow-up. The lowest for SDXL (M/Me 2.896/3.00) and DeepSeek-VL3 (M/Me 2.896/3.00), respectively. Mann-Whitney U test showed no significant differences between older and newer versions of models. The only noticeable difference between the architectures was the higher presence of stereotypes related to the use of the color blue and portraying people on the spectrum as loners, IT geeks, or artists in the case of the DALL-E model and the use of the child motif by diffusion models (see Fig.~\ref{fig:10boxplots}). 

The control prompts showed a statistically significantly lower degree of stereotyping non-autistic individuals with harmful autistic traits, with large size effects for all the models, thus confirming the hidden biases of the models. For DALL·E 3, the Mann-Whitney test yielded U = 478.50, Z = -6.63, p < 0.001, $\eta^2$  = 0.39; for Midjourney v7, U = 390.00, Z = -6.55, p < 0.001, $\eta^2$  = 0.41; for FLUX, U = 786.00, Z = -4.67, p < 0.001, $\eta^2$  = 0.20; for Stable Diffusion 3.5, U = 487.50, Z = -5.98, p < 0.001, $\eta^2$  = 0.34; and for DeepSeek-VL3, U = 332.00, Z = -6.91, p < 0.001, $\eta^2$  = 0.45 (see Table~\ref{tab:stereotyping} for overview of the differences between older-newer and experimental-control degree of stereotyping). 

The ratio of stereotypical themes to the number of images generated (baseline vs. follow-up) was found to be similar across the models (DALL-E: 2.91 vs. 3.53, Midjourney: 3.72 vs. 4.15, SDXL vs. FLUX: 2.90 vs. 2.93, Stable Diffusion: 3.92 vs. 2.74, Janus-Pro-7B: 3.36 vs. 3.28). This indicates that, in absolute values, a comparable degree of stereotyping was exhibited by both the DALL-E transformer architecture-based model, the models based on diffusion architecture, and the latest DeepSeek model. It is noteworthy that the proportion of males to females (281:86) depicted in the generated images closely resembles the proportion of genders in the clinical diagnoses. This is due to biases in diagnostic tools and procedures, which have resulted in autism being currently diagnosed 3 to 4 times more often in males. In this context, Janus appeared as the most “female-inclusive” model (61:26).

In addition to the three common stereotypes observed across all models (gender, skin color, and age), the most frequently repeated motifs for the models were (baseline/follow-up): DALL-E – the blue color theme / negative emotion; Midjourney – brain or modified head theme for both rounds; SDXL and FLUX – social isolation themes / blue color, Stable Diffusion – the color blue / negative affect, and Janus-Pro-7B – negative affect for both rounds. Among the three images with the highest degree of stereotyping at baseline (degree of stereotyping 8/10 and 7/10), two were generated by the Stable Diffusion model and one by Midjourney. In the follow-up (highest degree 7/10), two images were generated by Midjourney, one by Flux and one by Stable Diffusion. A notable distinction was the prevalence of stereotypes associated with using the color blue and the portrayal of individuals on the spectrum as isolated, nerds, or artists (in the case of the DALL-E model,) and the utilization of the child motif by diffusion models. 

In contrast, the degree of stereotyping in control images was noticeably lower in all models, except for the brain/modified head stereotype, which we found more often in the DALL-E and Stable Diffusion models. We explain this by the presence of the word “neurotypical” in the control prompts as an alternative to the word “autistic.” The dominant stereotypes of skin color and gender were significantly less present in all models, indicating that autistic people are largely identified with white men. In addition, differences in the degree of stereotyping of the categories “child” and “isolation” show that autistic people are more often than neurotypicals depicted as children and as people uninterested in social contact, which has incredibly serious social and clinical consequences. In the control set, the stereotype associated with the puzzle symbol (which was one of the more prominent in previous series) was almost absent. The puzzle motif did not appear at all in the DALL-E and Janus models (2x in Midjourney and 1x in Stable Diffusion and FLUX). Also, none of the images achieved as high a degree of stereotyping as in previous series (8/10 and 7/10), reaching a maximum of 5/10 (only 2 cases) and 4/10 (12 cases). 45 of the 265 control images did not contain any of the sought-after stereotypes, meaning that their level of autistic stereotypization was 0. 

\begin{figure}[h!]
    \centering
    
    \begin{subfigure}[t]{0.32\textwidth}
        \centering
        \includegraphics[width=\textwidth]{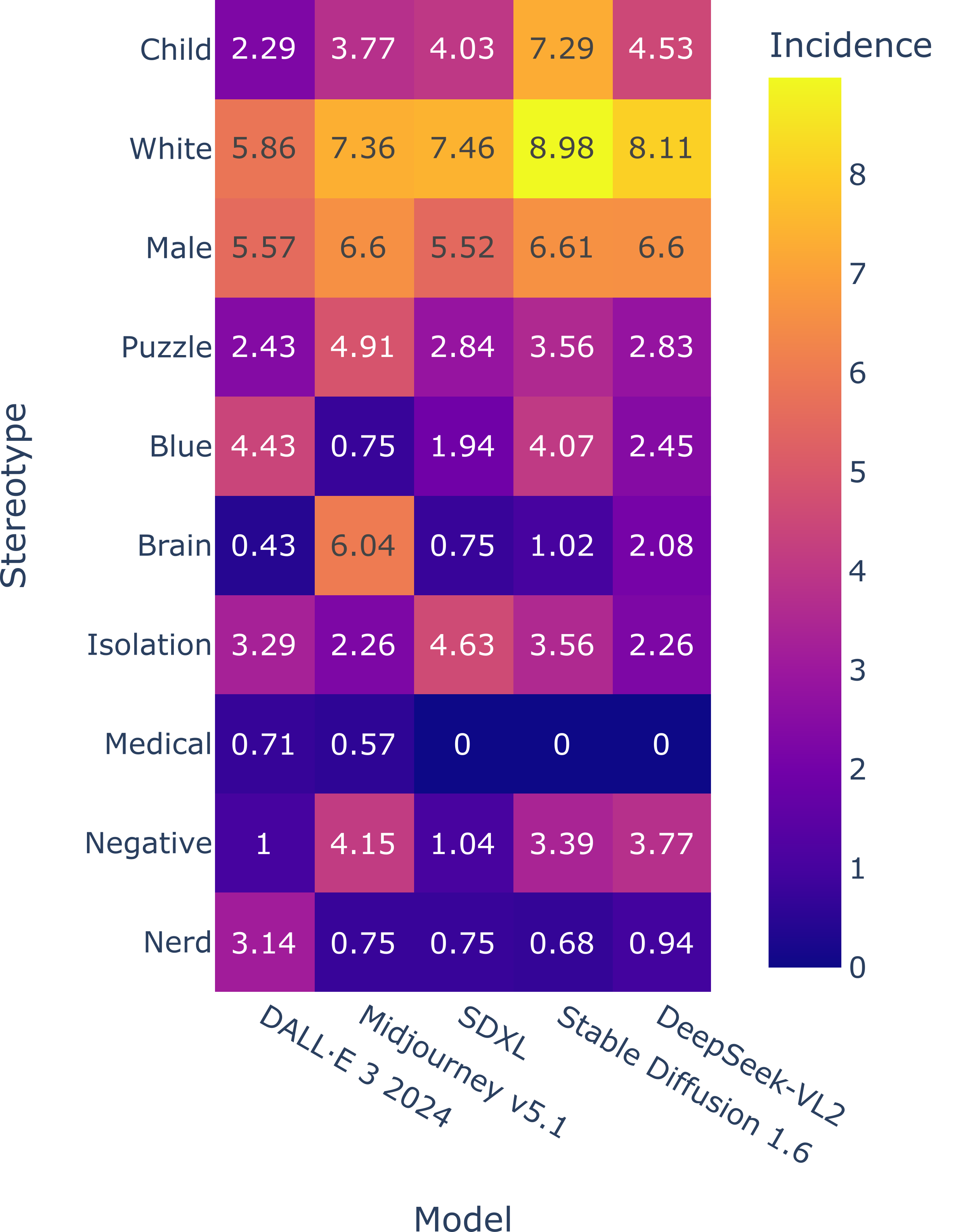}
        \caption{Proportion of average stereotype incidence across models - baseline}
        \label{fig:MAP}
    \end{subfigure}
    \hfill 
    \begin{subfigure}[t]{0.32\textwidth}
        \centering
        \includegraphics[width=\textwidth]{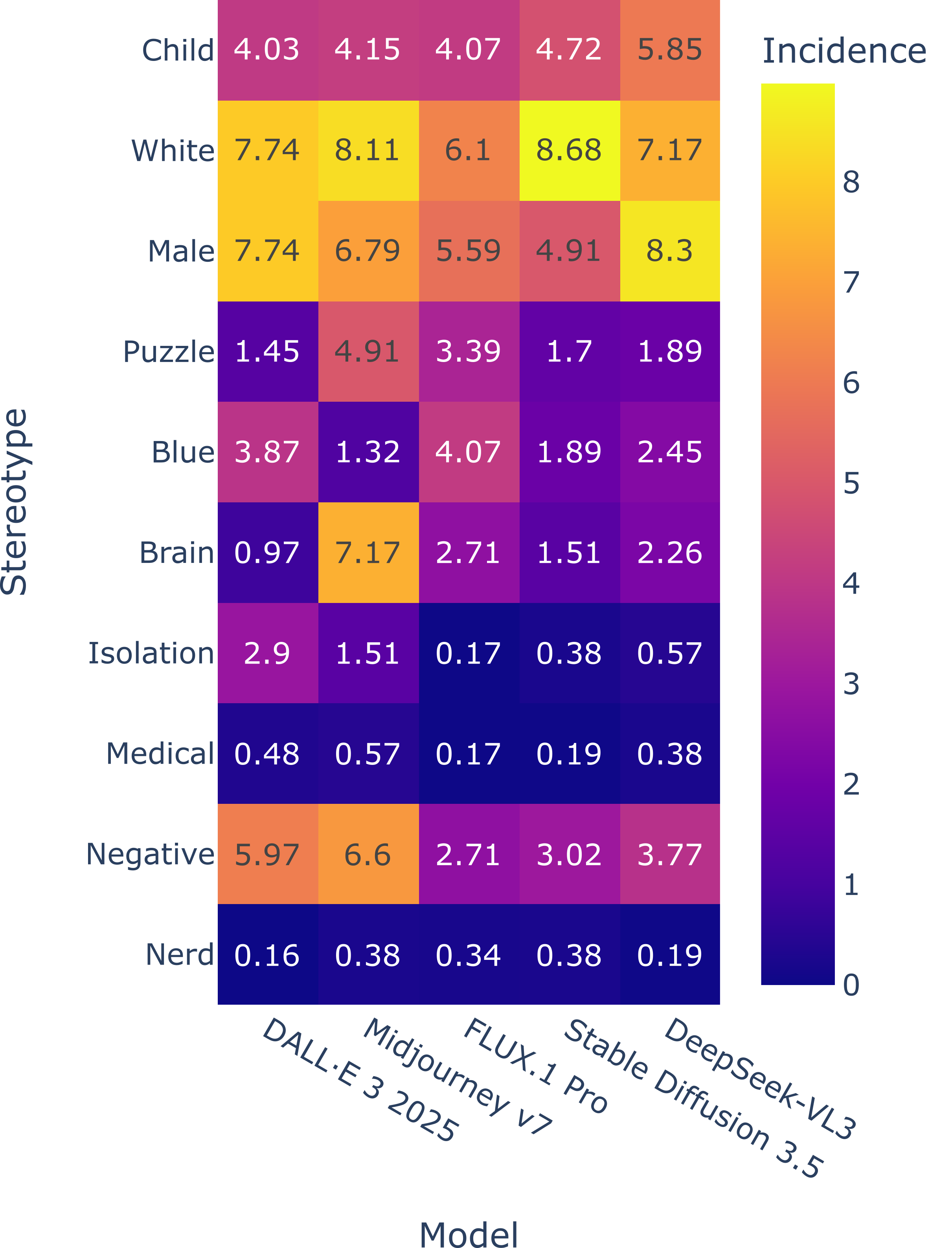}
        \caption{Proportion of average stereotype incidence across models - follow-up (2025)}
        \label{fig:heatmap2}
    \end{subfigure}
    \hfill 
    \begin{subfigure}[t]{0.32\textwidth}
        \centering
        \includegraphics[width=\textwidth]{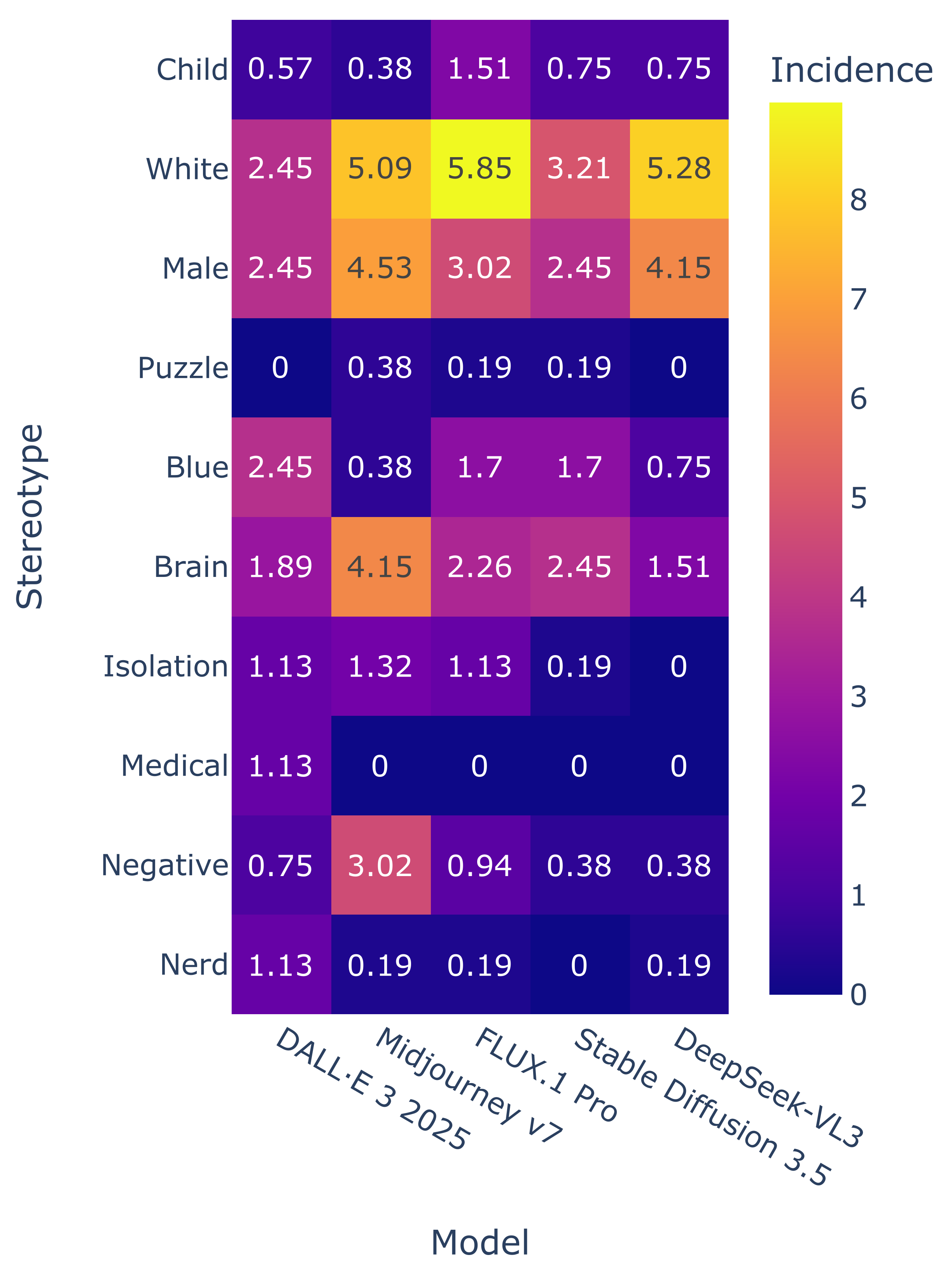}
        \caption{Proportion of average stereotype incidence across models - control images (2025)}
        \label{fig:heatmapkontrolna}
    \end{subfigure}
    
    \caption{Heatmaps showing the proportion of average stereotype incidence for baseline (a), follow-up (b), and control images (c)}
    \label{fig:heatmaps_all}
\end{figure}

Interestingly, stereotypes regarding the medicalization of autism were almost absent from the generated experimental graphics. This finding is intriguing in light of numerous contemporary analyses of media representations of autism, which have highlighted the prevalence and detrimental effects of portraying autism through a medical lens in media discourse \cite{Jones2023}, which apparently the AI models avoid. See Fig.~\ref{fig:MAP}, Fig.~\ref{fig:heatmap2}, and Fig.~\ref{fig:heatmapkontrolna} for the average incidence of the ten stereotypes for all models. The proportion of average stereotype incidence for each model is defined as the number of images in which a given stereotype was identified, divided by the total number of images generated by the model, and normalized by the maximum possible number of stereotypes (10). 

 \section{Discussion}
 \textbf{Recurring themes}\\
Regrettably, all the models perpetuated common stereotypes of autism. The most prevalent were: the white \citep{Brickhill2023} \citep{Cruz2024}, the young \citep{Aylward2021}, the boy \citep{Williams2022}, the puzzle symbol \citep{Gernsbacher2018}, and the blue color. The puzzle implies that autistic individuals are analogous to incomplete puzzles, lacking the components required for completion. This representation may influence the perception of autism as a deficit rather than as a diversity in human functioning; the symbol may also result in infantilization, whereby experiences and challenges are perceived as childish or trivial. The color blue has been criticized for its association with the controversial organization Autism Speaks and a perspective that focuses on males. This stereotype may contribute to the under-recognition and support for women and girls on the spectrum \citep{Cruz2024}. The prevalence of white male children among the depicted individuals serves to reinforce the erroneous assumption that autism is most prevalent in white boys. This one-sided representation results in the marginalization of the experiences of autistic people from different ethnic groups and cultures \citep{Williams2022}. Consequently, it can result in delays in diagnosis, inadequate support, and difficulties in accessing services for autistic individuals who do not align with this narrow perspective \citep{Mandell2009}. Also, a lack of diversity in representations can make it difficult to build social inclusion and understanding of the wide spectrum of autism in different communities.\\

\noindent \textbf{Group images}\\
Images of groups including or consisting of autistic individuals appear to be similar to each other and are less diverse than default groups without specified characteristics (see Fig.~\ref{fig:group_of_people} and Fig.~\ref{fig:controls_group_of_people}). In some instances, DALL-E generated explanations concerning the generated images, stating that the prompted issue is highly complex and that an alternative scene would be created in order to avoid perpetuating stereotypes (even though some of them were still used). It was evident from such cases that top-down ‘fairness protocols’ did not fully fulfill their role. Moreover, utilizing such safeguards does not address the underlying issue, which is the existence of biased datasets \citep{Norori2021}.\\

\begin{figure}[h!]
    \centering
    
    \begin{subfigure}[t]{0.48\textwidth}
        \centering
        \includegraphics[width=\textwidth]{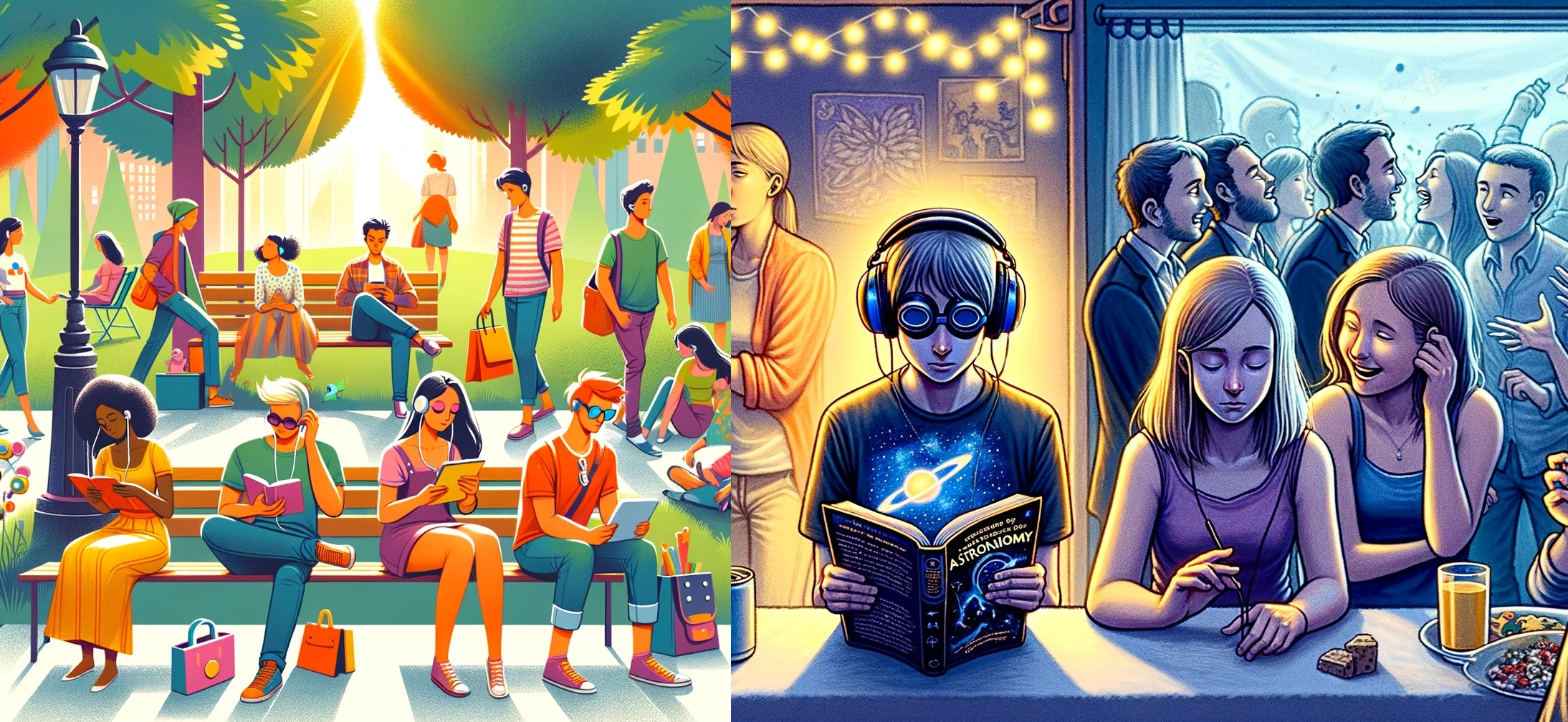}
        \caption{People on the spectrum depicted as focused on objects rather than personal relations. Prompt no. 9 (left) and 14 (right). Model: DALL-E (April 2024)}
        \label{fig:NOINT}
    \end{subfigure}
    \hfill
    \begin{subfigure}[t]{0.48\textwidth}
        \centering
        \includegraphics[width=\textwidth]{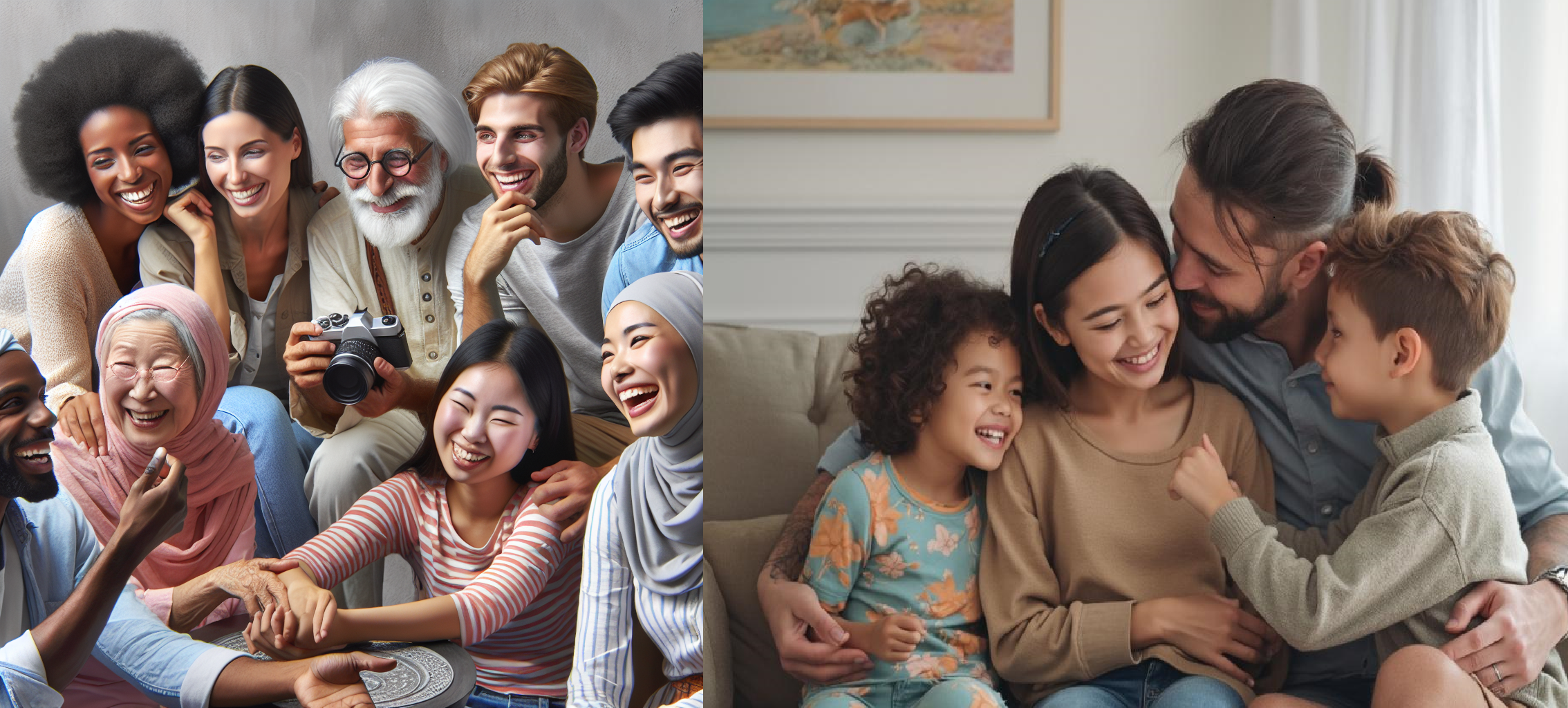}
        \caption{The people shown in the control images were more focused on interpersonal interactions. Prompt no. 9, DALL-E (August 2025) (left) and no. 25, FLUX (right).}
        \label{fig:controls_objects}
    \end{subfigure}
    
    \caption{Comparison of social focus in AI-generated images, contrasting object-focused depictions (a) with interpersonally-focused control images (b)}
    \label{fig:interaction_comparison}
\end{figure}

\noindent \textbf{Interaction: objects vs people}\\
 Images featuring multiple characters demonstrated a tendency to portray individuals on the autism spectrum as preoccupied with physical objects rather than engaging in interpersonal interactions. Even when these characters were in close proximity to one another, they did not engage in common activities, which reflects the pervasive (but often erroneous) belief that individuals with autism are antisocial, which is a hurtful stereotype \citep{Sinclaire2012} (see Fig.~\ref{fig:NOINT}). The people depicted in the control group's images were more often involved in interpersonal relationships. (see Fig.~\ref{fig:controls_objects})\\

\begin{figure}[h!]
    \centering
    
    \begin{subfigure}[t]{0.48\textwidth}
        \centering
        \includegraphics[width=\textwidth]{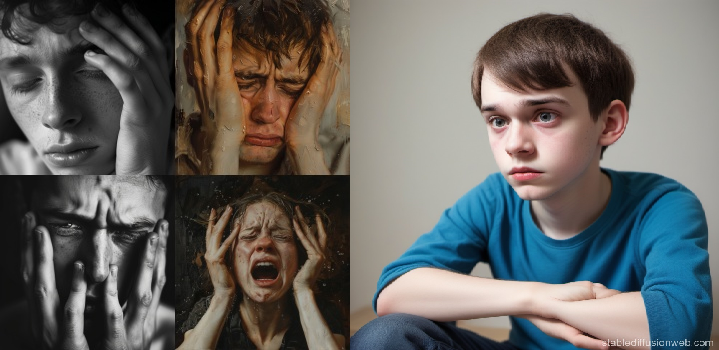}
        \caption{Difficult emotions and emotional blandness. Prompt no. 42, model: Midjourney v. 5.1 (left) and 33, model: Stable Diffusion v. 1.6 (right)}
        \label{fig:EMO}
    \end{subfigure}
    \hfill
    \begin{subfigure}[t]{0.48\textwidth}
        \centering
        \includegraphics[width=\textwidth]{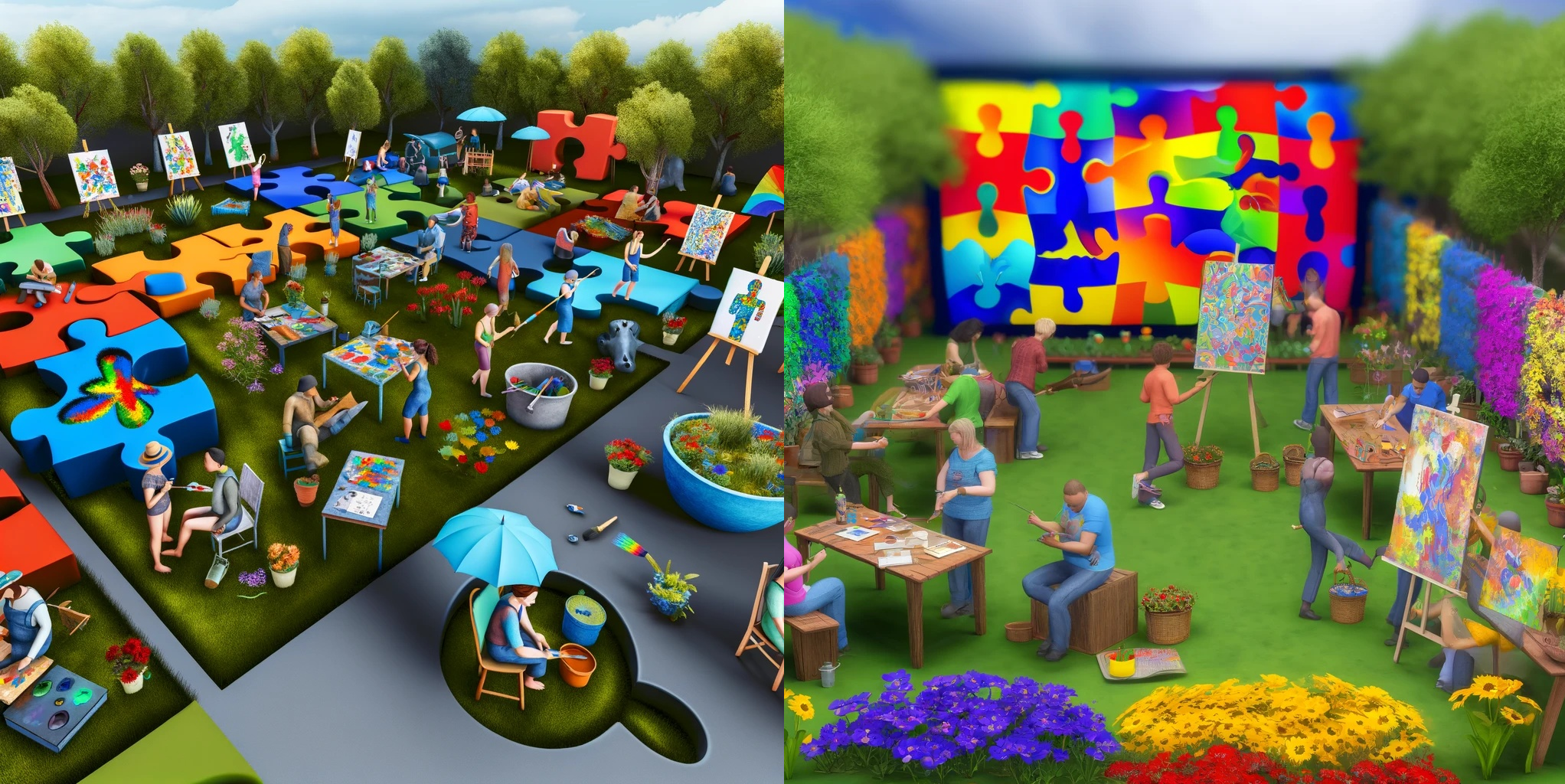}
        \caption{An image created with prompt no. 53: “Visualize autism without using a puzzle theme.”, model: DALL-E (April 2024)}
        \label{fig:NOPUZ}
    \end{subfigure}
    
    \caption{AI-generated visualizations of emotional states (a) and autism without the puzzle motif (b)}
    \label{fig:emotions_and_no_puzzle}
\end{figure}

\noindent \textbf{Emotional expressions and behavior}\\
In addition to the images in which the model was directly asked to display strong emotions, most of the characters presented in the generations exhibited a lack of emotional expressiveness. It is noteworthy that a greater number of images depicted positive emotions than negative ones. However, when the models were directly asked to generate images showing autistic people experiencing a strong emotion or showing a typical mood, the majority of images showed negative emotions. This may falsely suggest that autistic individuals do not experience intense emotions (or do so infrequently), or if they do, these are challenging, negative emotions rather than positive ones, such as joy or empathy \citep{Kimber2023} (see Fig.~\ref{fig:EMO}).

\vspace{1em}

\noindent \textbf{Artificial Neural Networks mirror human cognitive biases and mental imagery}\\
On a side note, we observed representational insensitivity regarding the generated images of autism despite directional prompting aimed at falsifying the stereotypes. For example, the most prevalent motif employed by models to represent autism was the puzzle symbol. Upon being explicitly instructed to generate visualizations that did not include this symbol, the models nevertheless incorporated it into their creations. The only model that handled this task properly was Janus. This could mean that either the model is better at tackling negative prompting, or it is better at distinguishing between the object classes it actually includes in the generated images. DALL-E explicitly refuted (on the text-to-text modality) the allegation that it perpetuates the puzzle stereotype, despite generating it. We hypothesize that insensitivity to negation may stem from encoder architecture, where embeddings of dominant tokens (e.g., “puzzle”) outweigh modifiers (e.g., “without”), causing cross-attention to preserve the dominant concept and ignore negation. This may also be interpreted as networks mirroring the human cognitive architecture regarding the discrepancy between background and reflective knowledge, as justified by research on autism-related stereotypes in humans. This analogy is grounded in psychological and neuroscientific research on implicit social cognition and stereotype activation, since artificial neural networks' statistical pattern completion may mirror the pattern of activation of entrenched cultural associations in human background knowledge. Furthermore, the images were frequently found to resemble the so-called human "mental images" (as different from "perceptual images") due to the presence of qualitatively undefined quantitative properties and a lack of adherence to the principle of individuation \citep{Beckmann2023}. This resulted, for example, in the simultaneous appearance of objects across multiple modalities. 
 
\begin{figure}[h!]
    \centering
    
    \begin{subfigure}[t]{0.48\textwidth}
        \centering
        \includegraphics[width=\textwidth]{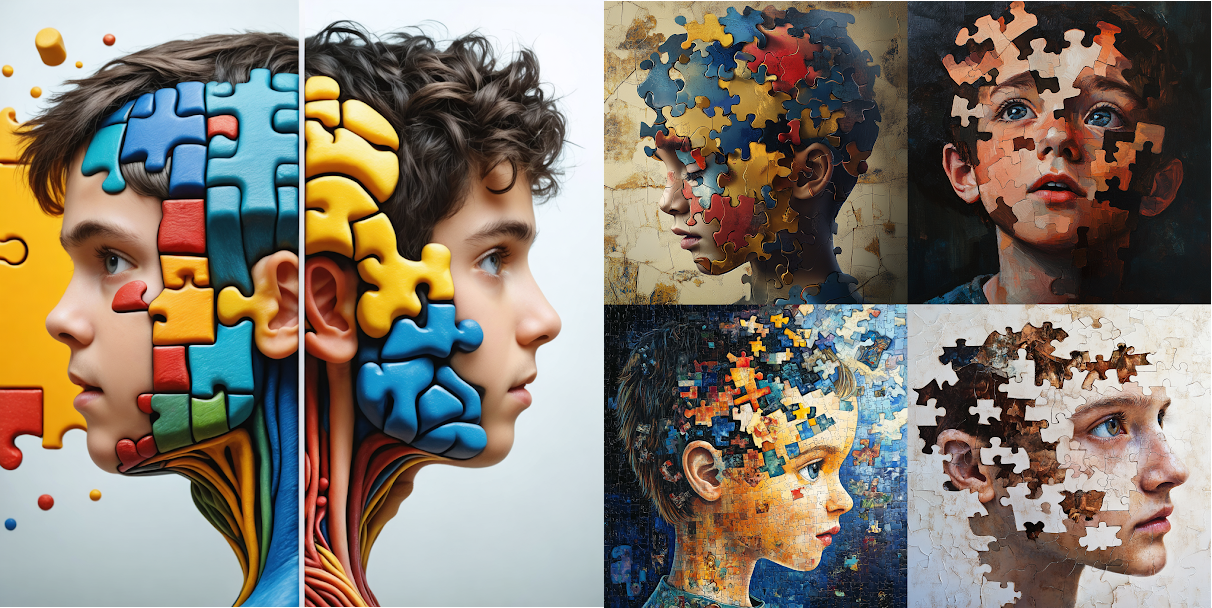}
        \caption{Examples of highly stereotyped images. Prompt no. 15, FLUX (left) and no. 1, Midjourney v. 7 (right).}
        \label{fig:fig11}
    \end{subfigure}
    \hfill
    \begin{subfigure}[t]{0.48\textwidth}
        \centering
        \includegraphics[width=\textwidth]{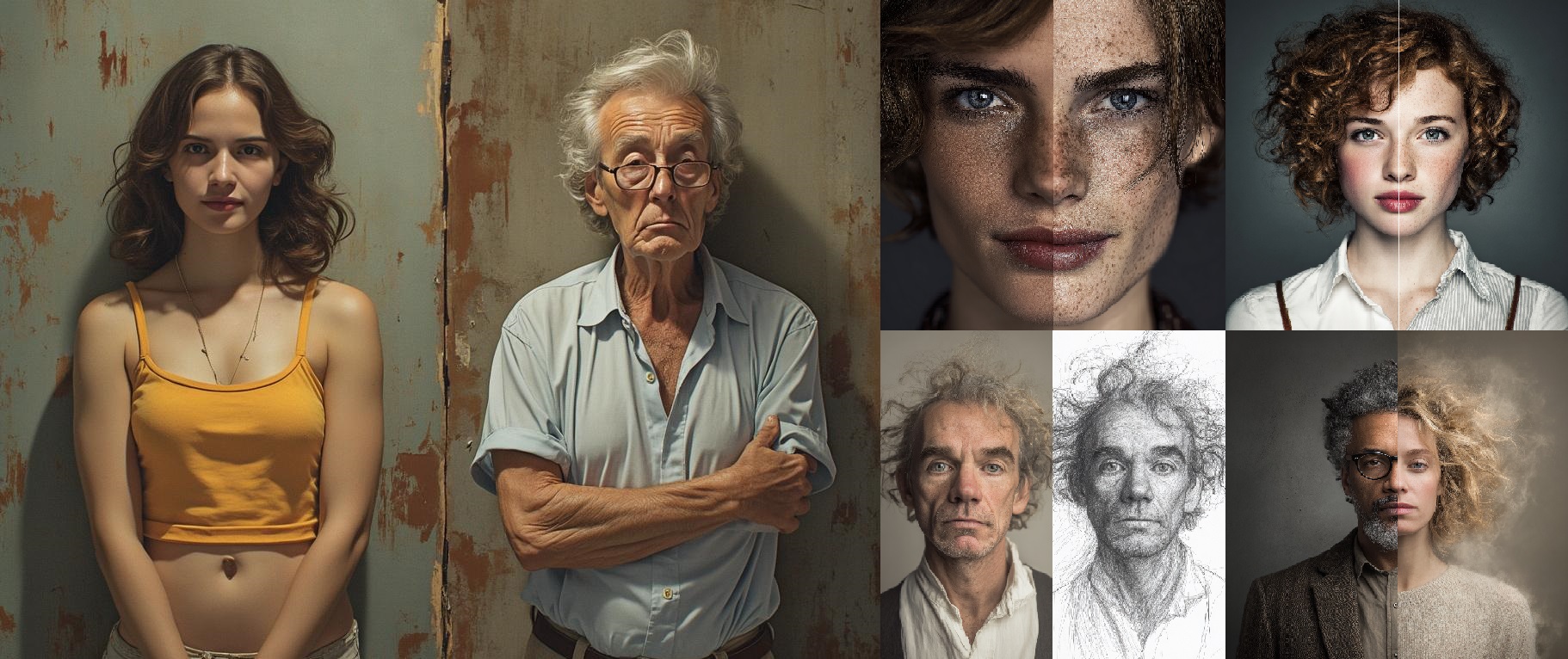}
        \caption{Control images for prompt 15. Models: Flux (left) and Midjurney v. 7}
        \label{fig:controls_15_15}
    \end{subfigure}
    
    \caption{Comparison of highly stereotyped images (a) with their corresponding control images (b)}
    \label{fig:stereo_vs_control}
\end{figure}

\noindent \textbf{Technological (computational) progress does not equal debiasing (ethical) progress}\\
This study shows that despite the undeniable technological advances that allow the models to generate images of higher quality and with fewer technical errors, the level of potentially harmful bias contained in the images remains largely similar. The aforementioned "low" median values of 3 and "high" of 4 in the 10-point scale are all actually high in absolute terms, given that the scale contains harmful stereotypes only, with particular images scoring as high as 7-8 on this scale (see Fig.~\ref{fig:fig11} and Fig.~\ref{fig:controls_15_15} for control images). In view of the intense discussions on the future of AI development and the place that ethics of AI aesthetics occupies, it should be underlined that models generating images are gaining importance also in terms of shaping the public epistemic structures. Top-down restrictions on the ability to generate visual content on certain topics or present a given aesthetic perspective will not solve the foundational issue arising from the inherent bias of the training data. In the end, our evaluation concerns the amplification and reinforcement of biases present in the human-created data by generative artificial intelligence, since pre-existing representations of autism created by humans are full of the analyzed stereotypes. The discussion of socially just and fair use of AI capabilities must also take this area into account. 

The question remains whether it is possible to create a good representation of an autistic person without using any stereotypes; in other words, whether such a person would be recognizable as autistic. In our view, autism often lacks visible features, and the expectation that AI-generated images “should” reveal visible traits only reinforces stereotypes. We suggest that the problem of generating stereotype-free images may not be simply difficult but perhaps structurally constrained. Models trained on biased data are unable to produce representations of autism that are both intelligible and free from stereotypes. It seems that "recognizability' itself is inherently tied to culturally shared but often reductive visual markers. Optimal data curation is rarely feasible, but biased generators can be partially corrected with post-hoc debiasing (e.g., concept erasure, model unlearning) and safety-oriented fine-tuning that penalizes stereotype-related activations during training \citep{Gandikota2023}. Future work could therefore examine whether multimodal models with deeper language understanding yield less stereotypical results from the same concise prompts. Nevertheless, we believe that AI models should not merely align with the majority of human-created data but strive for ethical alignment. If a model diverges from dominant biased patterns, reducing harm, this is a desirable outcome, not an error. Such deviations must be evaluated within interdisciplinary ethical frameworks, including neurodiverse ones, and not just statistical norms. In other words, models providing information not aligned with humanly created information may be “correct”. Identifying persisting harms is a necessary prerequisite for developing practical solutions, but ethical evaluation often lags behind technical innovation. Our results demonstrate that the analyzed models at their current stage of development may disseminate prevalent and harmful stereotypes regarding autism, and can thus be utilized as a repository of knowledge representing these stereotypes for research purposes. We express the hope that this work may contribute to the sensitivity regarding the appearances of neurodiverse individuals among LLM developers and, in the long run, serve the purpose of increasing the accountability of AI in the eyes of the autistic community. 

\pagebreak

\section*{Tables}
\vspace*{-10pt}
\renewcommand{\arraystretch}{1.2} 
\footnotesize
\begin{longtable}{p{1.5cm} p{6.5cm} p{8cm}}
    \multicolumn{3}{@{}l}{\small\sffamily\textbf{Table 1}} \label{tab:codes}\\
    \multicolumn{3}{@{}l}{\small\sffamily List of deductive codes and their descriptions} \\[1ex]
    \toprule
    \textbf{Code} & \textbf{Short Description} & \textbf{Explanation and Source} \\ 
    \midrule
    \endfirsthead 

    \toprule
    \textbf{Code} & \textbf{Short Description} & \textbf{Explanation and Source} \\ 
    \midrule
    \endhead 
        Child & Child (or the majority of the individuals are children, including teenagers) & The belief that autism concerns mainly children results in difficulties in diagnosing adults and the belief that autism can be ‘outgrown’ \citep{Aylward2021} \\

        White & White (or the majority of the individuals are white) & The racial stereotype of autism makes it difficult for non-white people to access diagnosis and treatment \citep{Cruz2024} \\

        Male & Male (or the majority of the individuals are males) & The false belief that autism primarily affects males is the result of bias in diagnostic tools and means that less attention is paid to symptoms occurring in women or non-binary people \citep{Williams2022} \\

        Puzzles & Puzzle theme somewhere in the picture (even if small or near the edge) & The symbol suggests that autistic people are ‘incomplete’. Such a metaphor may be seen as pejorative, suggesting that autism is a deficit rather than a difference in functioning \citep{Gernsbacher2018} \\

        Blue color & Blue is the dominant color (there is more blue than any other color; check even if blue accounts for less than 50\% of the total space or the central and most eye-catching object is blue) & Blue is often seen as a 'boy' colour, which can inadequately represent and marginalise females and non binary people on the autism spectrum. \\

        Modified brain, head & Modified brain/mind/head theme somewhere in the picture & Autism is ‘located’ in the head. Refers to the belief that autism is the result of a deficit located in the brain, especially when the image is cracked or falling apart \citep{Crawshaw2023} \\

        Isolation, loneliness & Themes of isolation (e.g., a person behind glass, closed doors, maze, cocoon, closed book, away from other people, loneliness), motifs and metaphors depicting hindered communication. Without interaction with other people & Stereotypically, people on the autism spectrum are perceived as devoid of empathy, unsympathetic, unwilling to establish contact, and socially isolated.\citep{Jones2009} \\

        Medical, disability themes & Medical / health-related / therapeutic / theme of disability (e.g., vaccinations, genes, intestines, wheelchair, walking cane, etc.). & Stereotype associated with the excessive medicalization of autism and the belief that it is something undesirable that needs to be ‘cured’ \citep{Wolbring2017} \\

        Negative emotions, emotional blandness, & Negative emotional state: sadness, upset, aggression, etc., or defragmentation (when the picture is, e.g, broken). Emotional blandness: face without expression, without emotion & People on the spectrum are often perceived as perpetually unhappy, “broken”, aggressive, and dangerous to those around them. \citep{Holton2014} \citep{Huws2011} \\

        Nerdy & Solitary Nerd, IT Geek, Scientist, etc. & Autistic people showed as lonely individuals, focused on unusual, complex interests that often require extraordinary abilities, which puts pressure on the majority of this social group who do not have such abilities \citep{Jordynn2014} \\

 \end{longtable}
 
 \begin{table}[h]
 	\centering
 	\caption{Degree of Stereotyping across models}
 	\label{tab:stereotyping}
    \renewcommand{\arraystretch}{1} 
    \footnotesize
 	\begin{tabular}{lccccc}
 		\toprule
 		\textbf{Model} & \textbf{U} & \textbf{Z} & \textbf{p} & \textbf{$\bm{\eta}^2$} & \textbf{Effect size} \\
 		\midrule
 		\multicolumn{6}{c}{\textbf{Baseline vs Follow-up}} \\
 		\midrule
 		DALL·E 2 vs DALL·E 3              & 1725.50 & -2.06 & $< 0.05$  & 0.03 & Small \\
 		Midjourney v5.1 vs Midjourney v7  & 1208.00 & -1.26 & $> 0.05$  & 0.06 & Small \\
 		SDXL vs FLUX.1 Pr                 & 1974.00 & -0.01 & $> 0.05$  & 0.00 & Negligible \\
 		Stable Diffusion 1.6 vs 3.5       &  900.00 & -3.94 & $< .001$  & 0.14 & Medium \\
 		DeepSeek-VL2 vs DeepSeek-VL3      & 1377.00 & -0.18 & $> 0.05$  & 0.00 & Negligible \\
 		\midrule
 		\multicolumn{6}{c}{\textbf{Experimental vs Control images}} \\
 		\midrule
 		DALL·E 3                          &  478.50 & -6.63 & $< 0.001$ & 0.39 & Large \\
 		Midjourney v7                     &  390.00 & -6.55 & $< 0.001$ & 0.41 & Large \\
 		FLUX.1 Pro                        &  786.00 & -4.67 & $< 0.001$ & 0.20 & Large \\
 		Stable Diffusion 3.5              &  487.50 & -5.98 & $< 0.001$ & 0.34 & Large \\
 		DeepSeek-VL3                      &  332.00 & -6.91 & $< 0.001$ & 0.45 & Large \\
 		\bottomrule
 	\end{tabular}
 \end{table}

\small
 

\section*{Declaration on Generative AI}
The authors have not employed any Generative AI tools except for the creation of research images. 
\bibliography{ECAI_2025_2}

\end{document}